\documentclass{amsart}

\numberwithin{equation}{section}

\usepackage{amssymb}
\usepackage{enumerate, xspace}
\usepackage{marvosym} 
\usepackage[sans]{dsfont}        
\usepackage{upgreek}
\usepackage{graphicx}
\usepackage[colorlinks]{hyperref}

\newtheorem{thm}{Theorem}[section]

\newtheorem{rem}[thm]{Remark}

\newcommand\cC{{\mathcal C}}

\newcommand\cE{{\mathcal E}}
\newcommand\cF{{\mathcal F}}

\newcommand\cL{{\mathcal L}}
\newcommand\cO{{\mathcal O}}
\newcommand\cM{{\mathcal M}}

\newcommand\bE{{\mathbb E}}
\newcommand\bN{{\mathbb N}}
\newcommand\bP{{\mathbb P}}

\newcommand\bR{{\mathbb R}}
\newcommand\bT{{\mathbb T}}
\newcommand\bX{{\mathbb X}}
\newcommand\bZ{{\mathbb Z}}
\newcommand\ba{{\boldsymbol{b}}}
\newcommand\bb{{\boldsymbol{a}}}

\newcommand\ve{\varepsilon}

\newcommand\vf{\varphi}

\newcommand\dive{{\operatorname{div}}}
\newcommand\Id{{\mathds{1}}}

\newcommand{\ce}{{\boldsymbol u}}
\newcommand{\cm}{{u}}
\newcommand{\bm}{\boldsymbol m}
\newcommand{\bi}{\boldsymbol i}
\newcommand{\id}{\operatorname{id}}

\newcommand\slow{z}
\newcommand\oTheta{{\bar\slow}}
\newcommand\sTheta{{\tilde\slow_\ve}}

\newcommand\shiftPar{\kappa}

\newcommand{\stdfSet}[2]{\bL_{#1}\ifx&#2&\else(#2)\fi}

\newcommand{\vei}{\ve^{-1}}
\newcommand{\veh}{\ve^{1/2}}
\newcommand{\Leb}{\textup{Leb}}
\newcommand\Var{\boldsymbol{\upsigma}}

\newcommand{\efrac}[2]{#1/#2} 

\newcommand{\nz}{n_Z}
\newcommand{\fpath}{h}
\newcommand{\dfpath}{\boldsymbol\partial \fpath}
\newcommand\intr{\textup{int}\,}
\newcommand\tpath[2]{$(#1,#2)$-\hspace{0pt}path}
\newcommand\admissiblep[2]{admissible \tpath{#1}{#2}}
\newcommand{\st}{\,:\,}

\newcommand{\deh}{\textup{d}}
\newcommand\bVar{\boldsymbol{\hat\upsigma}}

\begin{document}

\title{Transport in partially hyperbolic fast-slow systems}
\author{Carlangelo Liverani}
\address{Carlangelo Liverani\\
Dipartimento di Matematica\\
II Universit\`{a} di Roma (Tor Vergata)\\
Via della Ricerca Scientifica, 00133 Roma, Italy.}
\email{{\tt liverani@mat.uniroma2.it}}
\begin{abstract}
I will discuss, from a dynamical systems point of view, some recent attempts to rigorously derive the macroscopic laws of transport (e.g. the heat equation) from deterministic microscopic dynamics.
\end{abstract}
\thanks{Much if the work here described has been supported by the ERC grant  Macroscopic Laws and Dynamical Systems (MALADY) (ERC AdG 246953). The author would also like to thank Jean-Pierre Eckmann and Stefano Olla for pointing out several errors in a first version of this manuscript.}
\maketitle

\section{The problem}
In physics the world is described at different scales by seemingly very different laws. Once the laws are specified, the problem of explaining their compatibility in spite of their apparent differences becomes a mathematical one. Of course, it is possible to give heuristic explanations, and plenty of them are available. Nevertheless, it turns out that the issue is always very subtle, so that non rigorous explanations are often faulty and our naive intuition is at loss. 

In addition, in many instances one is interested in the behaviour of the world in a middle ground, that is at intermediate scales, and to make accurate predictions in such a realm a well grounded theory of how one scale merges in the next is necessary. An example of this kind is given by the current development of nanotechnology in which the systems of interest are mesoscopic: too large to apply easily the microscopic laws and too small for the macroscopic laws to be valid without qualification.

Here we will consider the transition between the macroscopic scale (the one we are used to) of the order of a meter and the microscopic (atomic) scale of the order of at most $10^{-9}$ meters.

While, ultimately, the cross over from the microscopic to the macroscopic must entail an understanding of the measurement process and of the semiclassical limit of Quantum Mechanics, many issues can be treated also remaining in the purely classical realm. One outstanding conceptual issue, going back to the ancient dispute between Zeno and Democritus 2400 yeas ago, stems from the fact that
the world around us looks like a continuum (we describe it using partial differential equations); yet we are aware that it consists of atoms, hence it is discrete in nature  (as first conclusively proven by Einstein \cite{Ei05} who showed how Brownian motion, a mesoscopic phenomena, emerges from the microscopic dynamics). 

It should therefore be possible to start from an atomic description and derive, in some appropriate limit that accounts for the difference in scales, a continuous description. In order to carry out such a program the first task is to identify the microscopic quantities that can be recognised and described macroscopically. It turns out that the microscopic dynamics has some quantities that are locally conserved (e.g. mass, energy, momentum, ...) and these tend to evolve much slower than the other degrees of freedom, hence allowing an evolution visible in the macroscopic time scale. In this article I will mainly discuss the situation in which the microscopic dynamics is Hamiltonian and classical and the conserved quantity is the energy. Hence the goal is to describe the energy evolution (energy transport) on the macroscopic scale.

Starting with the work of Boltzmann we understand that the microscopic energy manifests itself as thermal heat. The typical macroscopic law for heat transport is the Fourier law (although important violations of such a law, connected to specific microscopic properties, have been discovered, e.g. Carbonium nanotubes). 

For simplicity, we consider heat transport in homogenous non-conducting solids. This implies that we have a constant density and there is no mass flow: the only quantity that evolves is heat. The macroscopic definition of  heat is the amount of energy that is needed to change the temperature of a body and it is proportional to the temperature via the {\em specific heat per unit volume} $c_v(T)$.
The Fourier law states that the heat flux $J$ satisfies
\[
J=-\hat \kappa \nabla T
\]
where $T(x,t)$ is the temperature at point $x$ and time $t$, and $\hat\kappa$ is the {\em heat conductivity} of the material.  The assumption that heat is a locally conserved quantity is tantamount to saying that it satisfies a continuity equation:
\[
c_v(T)\partial_t T=-\dive J=\dive(\hat\kappa \nabla T).
\]
In other words, setting $\kappa(T)=\hat\kappa/c_v(T)$, often called {\em diffusivity}, we have
\begin{equation}\label{eq:heat}
\partial_t T=\kappa(T)\Delta T
\end{equation}
which is nothing else that the heat equation.

The mathematical problem mentioned above reads, in the present contest,   
\begin{itemize}
\item{\em Derive rigorously the heat equation from a Hamiltonian dynamics.}
\end{itemize}
 Such a problem is extremely hard, as is explained at length in the review \cite{BLR2000}. Such a  review is 17 years old, yet it is still actual since, in spite of considerable efforts, little progress has been achieved in the intervening years. Yet, little progress is not zero progress. In the following I will describe some of the mathematical work made in the last years. I start by making precise the kind of systems I want to consider.
 
\section{The models: general considerations}
We describe here a very idealised model of a homogeneous, non-conducting solid. In other words, the solid will look more or less the same at all places and the particles that constitute it are not free to move very far from their rest position. This is the simplest possible situation one can think of since the only quantity that can move around is the energy.

The main feature of the problem is that there is a microscopic versus a macroscopic world (and description). As we will restrict ourselves to the classical world (that is, we are ignoring quantum effects) both world can be described by differential equations (either ordinary, stochastic or partial) on $\bR^n$, for some $n\in\bN$. In fact, the macroscopic world and the microscopic one differ just in the scale. The distinction can be encapsulated in the fact that a scaling parameter $L$ becomes extremely large.

More precisely, since we are going to consider only models of solids, we can restrict  our discussion to microscopic models defined in a region $\Lambda_L=\{x\in\bZ^d\;|\; L^{-1}x\in\Lambda\}$ for some nice, fixed, region $\Lambda\subset \bR^d$.\footnote{ Of course we are mainly interested in the cases $d=1,2,3$ (wires, membranes and solids).} Accordingly, the region $\Lambda$ stands for the macroscopic solid we want to consider, while $\Lambda_L$ is the corresponding region in the microscopic description. Note that $\Lambda_L$ is discrete in nature as it is a subset of a lattice and it has size of order $L$ when measured in microscopic units.\footnote{ The choice of a square lattice is immaterial, any other will do.} Note that the microscopic unitis are such that the discrete nature of the system is evident at distances of order one, rather than at distance $L^{-1}$, as it happens in macroscopic units.

At each point $x\in\Lambda_L$ we assume that there is a group of particles (atoms, molecules, defects, ...) that are described by coordinates $q_x,p_x\in M_*$,\footnote{ The fact that $M_*$ does not depend on $x$ is part of our homogeneity hypothesis.} where $M_*$ is the cotangent bundle of an $n_*$ dimensional compact Riemannian manifold $M$. One can think of the $q_x$ as the displacement of particles from their equilibrium positions while the $p_x$ are their momenta (or velocities). The fact that $q_x$ belongs to a compact manifold is our non conduction hypotheses: particles are not free to move around the solid. That is, there is no {\em convection}.\footnote{ Of course, one can achieve the same when $M_*=\bR^{2n_*}$ by introducing some strongly confining potential, see section \ref{sec:othermodels}, but for now let us keep things as simple as possible.} Accordingly, the microscopic system is described in a phase space $\cM_L= M_*^{\Lambda_L}$.

On the contrary the macroscopic description consists simply of a temperature field $T(x,t)$, $x\in\Lambda$, $t\in\bR$. Since the temperature is a function of the internal energy density $\cm$ of the body, the heat equation \eqref{eq:heat} can be written as
\begin{equation}\label{eq:heate}
\partial_t \cm=\dive(\kappa(\cm)\nabla \cm).
\end{equation}

Equation \eqref{eq:heate} describes the macroscopic dynamics, it remains to describe the microscopic dynamics. Ideally we want a Hamiltonian dynamics, yet it is instructive to allow also stochastic dynamics, since their study is easier and it might provide important ideas to understand the deterministic case.
Nevertheless, we will assume that the dynamics is Markov. In other words, given any initial probability distribution $\bP_0$,  the distribution $\bP_t$ at time $t$ is given by, for all $f\in\cC^0(\cM,\bR)$,
\begin{equation}\label{eq:semig}
\bE_t(f)=\bE_0(\cL_t f)
\end{equation}
where $\bE_t$, $t\geq 0$, is the expectation with respect to the probability measure $\bP_t$ and $\cL_t: \cC^0\to\cC^0$ is a strongly continuous one parameter semigroup. Of course, in the case of a Hamiltonian dynamics, calling $\phi_t$ the Hamiltonian flow, we will have $\cL_t f=f\circ \phi_t$.

To specify the dynamics we need to discuss also the initial conditions. While the initial conditions of equation \eqref{eq:heate} are simply an initial energy profile $u_0(x)$, the initial conditions for the microscopic model are a much more subtle issue, especially in the case of a Hamiltonian dynamics. Indeed, Hamiltonian dynamics are reversible: for each trajectory there exists an initial condition for which the trajectory is run backward (just take the final configuration of the trajectory and reverse all the velocities). On the contrary the heat equation \eqref{eq:heate} has no such property. So initial conditions must play an important role. To address this problem we must  be more specific about the type of dynamics we consider, so we postpone the discussion momentarily (see \eqref{eq:inco} for details). For the time being we require the bare minimum. First of all we consider random initial conditions, that is the initial conditions are described by a {\em non atomic} measure $\bP_0$. This corresponds to the natural fact that the exact positions and velocities of the microscopic particles (that in real life applications may be of order $10^{23}$) cannot be known precisely and only a statistical knowledge is possible. 

Also, to connect to the macroscopic setting, we must say what we mean for the internal energy $\ce_x$ of the body at site $x$. This can be done in several ways that are all essentially equivalent, however they depend on the form of the dynamics which we have not yet described precisely; so we postpone the definition to equation \eqref{eq:energyx}. However, whatever the exact definition, we  are interested in the measures $\mu_{\ce,L}$ defined as, for all $\vf\in\cC^\infty$, \footnote{ To simplify matter we assume that the volume of $\Lambda$ is one.}
\[
\mu_{\ce,L}(\vf):=\frac 1{L^d}\sum_{x\in\Lambda_L}\vf(L^{-1} x)\ce_x(0)=\frac 1{L^d}\sum_{x\in\Lambda_L}\ce_x(0)\delta_{L^{-1} x}(\vf).
\]
Note that $\mu_{\ce,L}$ depends from the microscopic configurations and hence is a random variable under $\bP_0$. We then ask that there exists a smooth function $\cm$ such that
\[
\lim_{L\to\infty}\frac 1{L^d}\sum_{x\in\Lambda_L}\ce_x(0) \delta_{L^{-1} x}(\vf)=\int_{\bR^d} \cm(x)\vf(x) dx
\]
where the limits is meant almost surely with respect to $\bP_0$. Thus, at time zero the microscopic energy gives rise to a nice energy profile on the macroscopic scale.

We are finally able to specify in which sense the macroscopic dynamics should arise from the microscopic one:  for each $\vf\in\cC^\infty(\bR^d,\bR)$, consider the measures
\[
\mu_{\ce,L,t}(\vf):=\frac 1{L^d}\sum_{x\in\Lambda_L}\vf(L^{-1} x)\ce_x(L^2 t)=\frac 1{L^d}\sum_{x\in\Lambda_L}\ce_x(L^2 t)\delta_{L^{-1} x}(\vf).
\]
The measure $\mu_{\ce,L,t}$ describes the energy density in the microscopic system at the microscopic time $L^2 t$. The choice for this time scaling (called {\em parabolic} or {\em diffusive} scaling) comes from the fact that equation \eqref{eq:heate} is invariant under such a scaling, so it presents itself as the natural one. If the macroscopic dynamics must arise from the microscopic dynamics, the we expect that $\bP_0$-a.s.
\[
\lim_{L\to\infty}\frac 1{L^d}\sum_{x\in\Lambda_L}\ce_x(L^2 t)\delta_{L^{-1} x}(\vf)=\int_{\bR^d} \cm(x,t)\vf(x) dx,
\]
where $\cm$ satisfies \eqref{eq:heate}. 

It turns out that the above limit is hard to justify even at the heuristic level, so, as a preliminary step, one would be rather happy even proving its averaged version:\footnote{ Note that here we abuse notation and use $\bP_0, \bE_0$ to designate, respectively, the measure and expectation in path space determined by the initial measure (that we also called $\bP_0$, hence the abuse). Of course, in the deterministic case all is determined by the initial condition, but in the random case the measure in path space describes also the randomness of the dynamics.}
\begin{equation}\label{eq:hydro}
\lim_{L\to\infty}\frac 1{L^d}\sum_{x\in\Lambda_L}\bE_0(\ce_x(L^2 t)\delta_{L^{-1} x}(\vf))=\int_{\bR^d} \cm(x,t)\vf(x) dx.
\end{equation}
This type of results are called {\em hydrodynamic limits} and have been first obtained in some generality  in the context of stochastic microscopic dynamics, \cite{KMP, dipp, GPV, Va, ovy}. For an overview of the hydrodynamic limit see \cite{Sp}.

I have thus specified what can be considered a satisfactory explanation of the emergence of the macroscopic dynamics \eqref{eq:heate} from a microscopic model. Of course, this, rather than being the end of the story, is just a starting point. In fact, what are really relevant for physics and applications are the {\em finite size effects}. That is, the corrections to the macroscopic law coming form the fact that the scale difference ($L$) is finite and not infinite. This are the type of results that could prove relevant when working at the mesoscopic scale (e.g, nanotechnology).

\section{The models: microscopic dynamics}
To make precise the model we have to specify the dynamics. Let us start from a Hamiltonian dynamics: this is the one physicists would ultimately like to study. 

By our simplifying homegeneity hypothesis the local systems have all the same local Hamiltonians  
\begin{equation}\label{eq:localh}
h(q,p)=\frac 1{2}\langle p,\bm^{-1} p\rangle+U(q)
\end{equation}
for some strictly positive matrix $\bm$ and smooth potential $U$. To simplify notation we assume $\bm=\Id$.\footnote{ Note that we can always reduce to this situation by changing the definition of the scalar product.} The global Hamiltonian is the sum of the local Hamiltonians and of the interaction between near by systems. For simplicity again we assume that the interaction takes place only among nearest neighbors. Also, since we are considering the system as a bunch of interacting systems, we expect the typical internal energy of a local system (binding energy) to be much larger than the interaction energy. We are thus led to a global Hamiltonian of the form
\begin{equation}\label{eq:ham-tot}
H_{\ve,L}(q,p)=\sum_{x\in\Lambda_L} h(q_x,p_x)+\frac \ve 2\sum_{x\in\Lambda_L}\sum_{\|y-x\|=1}V(q_x,q_y)
\end{equation}
where $V(q,q')=V(q',q)$ is a symmetric smooth potential.

We can now specify what we mean by the internal energy at site $x$:
\begin{equation}\label{eq:energyx}
\ce_x=h(q_x,p_x)+\frac \ve 2\sum_{\|y-x\|=1}V(q_x,q_y).
\end{equation}
We added to the local Hamiltonian the interaction energy so that $\sum_x\ce_x=H_{\ve,L}$, thus the $\ce_x$ account for all the energy in the system.

\begin{rem}\label{rem:bc} Note that we are considering the case of a body in isolation. In reality the bodies are in contact with the exterior that can be thought as a thermal reservoir at some given temperature. This is, of course, an extremely important problem but it has several extra difficulties (for example, one has to decide a model for the thermal reservoir, and this is the subject of many debates; the invariant measure of the dynamics is not known explicitly, and even establishing its existence is a challenge, see section \ref{sec:othermodels} for more details). Accordingly, to keep the exposition simple, we will not discuss boundary conditions and we will only consider isolated bodies.
\end{rem}
Let $\phi_t$ be the flow generated by \eqref{eq:ham-tot} via the usual Hamilton equations
\begin{equation}\label{eq:hameq}
\begin{split}
\dot q&=\partial_p H_{\ve,L}\\
\dot p&=-\partial_q H_{\ve,L}.
\end{split}
\end{equation}
We have already mentioned that the semigroup \eqref{eq:semig} is defined as $\cL_t f=f\circ \phi_t$. A simple computation shows that the generator of $\cL_t$ is given by
\[
\begin{split}
\frac{d}{dt}\cL_t f|_{t=0}=:\bX f=&\sum_{x\in\Lambda_L}\langle p_x,\partial_{q_x}f\rangle -\langle\nabla U(q_x), \partial_{p_x} f\rangle\\
&-\ve\sum_{x\in\Lambda_L}\sum_{\|y-x\|=1}\langle \partial_{q_x}V(q_x,q_y), \partial_{p_x} f\rangle,
\end{split}
\]
where we have used the symmetry of $V$. 

The first problem in tackling the above dynamics is that we are interested in the properties of the system for a very long time (of order $L^2$). At the moment the only dynamics that are well understood for arbitrary long times are: a) completely integrable systems; b) strongly chaotic systems. The first possibility is of course much simpler, unfortunately it is very non generic. The interaction between different local systems will typically break the complete integrability leading to a system that we have no tools to analyse. 

Of course, one could consider very special global systems that are completely integrable, for example a system in which all the potentials are quadratic (harmonic systems), leading to linear Hamilton equations \eqref{eq:hameq}. Indeed the exploration of such systems started a long time ago \cite{RLL67, LLL} but it yields an anomalous diffusion due to the existence of many conserved quantities beside the energy \cite{Zo02}. This goes against the general consensus that the only locally conserved quantity should be the energy. To ensure such a fact one can introduce some stochasticity in the system (either in the interactions or in the local dynamics) and indeed several very interesting results have been obtained concerning harmonics crystals with some randomness, see \cite{OS13, BO14, JKO15, KO16, KO17} or the review \cite{BBJKO} and references therein.

In general, the introduction of noise makes the problem much more tractable. If the noise is sufficiently strong, then it is possible to establish the full hydrodynamic limit \cite{ovy, LO96, FLO97}, but also for a very degenerate noise relevant partial results can be obtained \cite{LO12}. 

The alternative is to consider strongly chaotic local dynamics. This point of view has been first advocated in a precise manner by \cite{GG08}, in which they propose to study a billiard type model inspired by \cite{BLPS}.\footnote{ Billiards do not fall in the class of models described by \eqref{eq:ham-tot} because the potential is not smooth as they have hard-core interactions. However, they are Hamiltonian and morally similar. We will comment further on hard core models in section \ref{sec:hard}.} This is the point of view I wish to pursue: assume that the Hamiltonian flow associated to the Hamiltonian $h$ is strongly chaotic (an {\em Anosov} flow with {\em exponential decay of correlations}).

Nevertheless, as we mentioned already, the introduction of a stochastic part in the dynamics is very instructive. To illustrate this we will consider the case in which the interaction between nearby systems has a very strong random component.  We will see that this can be partially justified as a mesoscopic regime, see Theorem \ref{thm:dolgo}, but for the time being it is just an heuristic tool. We assume that near by systems exchange their velocities and that on each kinetic energy surface takes place a diffusion. To make the statement precise, consider the vector fields
\[
\begin{split}
Y_{i,x}f(q,p)&=\langle y_i(p_x), \partial_{p_x}f(q,p)\rangle\\
X_{x,y}f&=p_x\partial_{p_y}f-p_y\partial_{p_x} f
\end{split}
\]
where the vectors $\{y_i(p)\}$ spans the tangent space of the kinetic energy surface $\{\bar p\in\bR^d\::\;\bar p^2=p^2\}$ at the point $p$.
We then consider the semigroup generated by 
\begin{equation}\label{eq:generats}
\begin{split}
&\bX_\nu=(1-\nu) \bX+ \nu S\\
&S=\sum_{x\in\Lambda_L}\sum_i Y_{i,x}^2+\sum_{x\in\Lambda_L}\sum_{\|x-y\|=1}X_{x,y}^2.
\end{split}
\end{equation}
Let $\cL_{\nu,t}$ be the semigroup generated by $\bX_\nu$. Note that $\cL_{0,t}=\cL_t$ is the deterministic dynamics, while $\cL_{1,t}$ is a purely stochastic (diffusive) dynamics that does not move the $q$ and preserves the kinetic energy $\sum_x  p_x^2$. Thus, it reduces to a purely momenta dynamics in which the positions do not play any role.

\section{Invariant measures, reversibility and currents}
It is well known that a Hamiltonian flow leaves invariant the Liouville measure $m_{L,E}$, that is the uniform measure on the energy surface $\cM_{L,E}$.
This implies that the total energy is an invariant quantity for the generators $\cL_{\nu,t}$ and the Liouville measures $m_{L,E}$ are invariant. 
Moreover, for $\ve=0$, all the Liouville measures $m_{L, \overline{E}}$, $\overline{E}=(E_x)$, supported on the energy surfaces $\cM_{L, \overline{E}}=\{(q,p)\in \cM_{L,E}\;:\; \ce_x=E_x, \sum_x E_x=E\}$, are invariant.

A satisfactory class of initial measures $\bP_0$ we may wish to consider is given by
\begin{equation}\label{eq:inco}
\bE_0(f)=\rho(p,q)m_{L, \overline{E}^*}
\end{equation}
for some smooth integrable function $\rho$ and energies $\sum_x E^*_x=E$. That is, we are allowed to fix the energies (slow variables) but not the fast variables.

Note that
\[
m_{L,E}(f Sf)=-\sum_{x\in\Lambda_L}\sum_i m_{L,E}((Y_{i,x} f)^2)-\sum_{x\in\Lambda_L}\sum_{\|x-y\|=1}((X_{x,y}f)^2),
\]
which implies
\begin{equation}\label{eq:rev}
m_{L,E}(f \cdot Sf)=m_{L,E}( Sf\cdot f).
\end{equation}
On the contrary, since $m_{L,E}(f\cdot  f\circ \phi_t)=m_{L,E}(f\circ \phi_{-t}  f)$, we have
\begin{equation}\label{eq:anti}
m_{L,E}(f \bX f)=-m_{L,E}(f \bX f).
\end{equation}

A semigroup, with the property \eqref{eq:rev} (that is, its generator is self-adjoint), is called {\em reversible}.

Reversibility is an important property for Markov systems, which has many relevant consequences \cite{KV86}. Unfortunately, the generator of  a deterministic system is anti-selfadjoint (see \eqref{eq:anti}), which is as far as possible from reversible. 

However, there exists a seemingly rather different definition of reversibility in the context of flows. A definition which also has momentous consequences \cite{GC95}. 
We call a flow $\phi_t$ {\em reversible} if there exists an involution $\bi:\cM_L\to\cM_L$, that is $\bi\circ\bi=\id$,
such that
\[
\phi_t\circ \bi=\bi\circ \phi_{-t}.
\]
In the case of a Hamiltonian system it is trivial to check that the flow is reversible with the involution $\bi(q,p)=(q,-p)$.
Note that the macroscopic evolution associated to the equation \eqref{eq:heate} can also be seen as a semigroup where the generator is $Af=\dive(\kappa(f)\nabla f)$.\footnote{ Note that the \eqref{eq:heate} is really the equation for the density of a measure, so technically the semigroup here is the adjoint of the ones we were discussing above.}  However such equation is {\em irreversible}, where the word refers to the fact that backward dynamics is not well defined.\footnote{ In a sense the equation is irreversible if considered a deterministic equation (which is its physical meaning). It is instead reversible, at least in the case $\kappa(f)=\kappa$, if the associated semigroup is interpreted as the semigroup describing a random process (Brownian motion). Sorry for the ambiguity.} Thus the two different definitions of reversibility are yet another manifestation of a long standing conundrum: the relation between microscopic reversibility and macroscopic irreversibility (or: how to explain the time arrow).

We are now ready to discuss another important issue: the {\em current}. The current simply describes the change in energy of a local system. A direct computation yields, for all $s<t$,
\[
\begin{split}
&\frac{d}{dt}\bE_0(\ce_x(t)\;|\; \cF_s)=\bE_0(\bX_\nu \ce_x(t)\;|\; \cF_s)= \sum_{\|y-x\|=1} \bE_0(j_{x, y}(t)\;|\; \cF_s)\\
&j_{x,y}=(1-\nu)\frac{\ve} 2\left[p_y\partial_{q_y}V(q_x,q_y)-p_x\partial_{q_x}V(q_x,p_y)\right]+\nu (p_y^2-p_x^2),
\end{split}
\]
where $\cF_s$ is the $\sigma$-algebra determined by the variables $\{q(\tau),p(\tau)\}_{\tau\leq s}$. Note that $j_{x,y}=-j_{y,x}$, so the total energy is conserved.

In the case $\nu=1$, since the $q$ do not evolve, the energies $u_x(t)$ differ from $p_x^2(t)$ only by a constant, so we get a closed equation for the kinetic energy
\[
\frac{d}{dt}\bE_0(p_x^2(t))= \sum_{\|y-x\|=1}\bE_0(p_y^2(t))-\bE_0(p_x^2(t)).
\]
in this case the current is an exact discrete gradient. Also it is not hard to prove that the measures $\mu_{p^2,L,t}$ are tight, so for any convergent subsequence we have, for each $\vf\in\cC^\infty$,
\[
\mu_{p^2,L_j, t}(\vf)-\mu_{p^2,L,0}(\vf)=\int_0^{t} \mu_{p^2,L_j, t}(\Delta \vf)+\cO(L_j^{-1}).
\]
The above, using the notation of \eqref{eq:hydro}, yields
\[
\partial_t u(x,t)=\Delta u(x,t)
\]
in the weak sense. This is an extreme manifestation of the fact that gradient currents are easier to treat since a Laplacian is already implicit in the current. See \cite{GPV} to see how to treat the general gradient case. When the current is not of a gradient type (as in the case $\nu\not =1$), then much more work is needed, see \cite{Va}.

The standard tools to deal with the non-gradient case (say $\nu=0$) seem to require two facts: 
\begin{enumerate}[a)]
\item the well posedness of the Green-Kubo formula
\begin{equation}
  \label{eq:gk1}
  \kappa_\ve= \beta^2 \ve^2 \int_0^\infty \sum_{x\in \bZ^d}  \bE_{\beta} \left( j_{x,x+1}(t) j_{0,1}(0) \right) dt,
\end{equation}
for the infinite system at equilibrium with inverse temperature $\beta$. 
\item A spectral gap of order $L^{-2}$ for the dynamics $\cL_t$ in a region of size $L$.
\end{enumerate}
 Of course, the latter refers to stochastic systems where the gap is meant in $L^2$ or in some simple Sobolev space. In the deterministic case on such spaces there is no gap at all. Yet, there can be exponential decay of correlations for smooth observables, Accordingly, it is  likely that any possible proof will require a decay of correlations of type $e^{-C L^{-2}t}$ for  reasonable observables and the dynamics in a region of size $L$.

Accordingly, all the known approaches require a sharper, quantitative, information on the rate of convergence in the formula \eqref{eq:gk1}.
Note that, even assuming that the flow on each energy surface of the local dynamics is Anosov, already the study of two interacting systems is currently out of reach. Indeed, when two systems interact only the total energy (and not the individual ones) is conserved. Hence, two interacting systems can be viewed as a {\em partially hyperbolic flow} with a three dimensional central direction. No result whatsoever is currently available on the rate of mixing for such systems, let alone a larger collection of interacting systems. See \cite{BDV05} for an overview on partially hyperbolic systems.

\subsection{Other models}\label{sec:othermodels}\ \newline
Let us briefly discuss other possibile microscopic models. One possibility we already mentioned is that $M_*=\bR^{2n_*}$ but the confinement is provided by a potential, these are essentially anharmonic chains (we have already mentioned the harmonic case). The first example of such a model goes back to Fermi-Pasta-Ulam (FPU) \cite{FPU}. The FPU models have proven extremely difficult to investigate, even numerically \cite{BLP, BPC, DPS}. However, the study of FPU models has shown that the route from microscopic to macroscopic is much subtler than one can naively imagine and metastable states may play an important role. From the rigorous point of view almost nothing is known, apart from some zero energy density results that are not so relevant in the present context. On the other hand, if one considers the case in which the system is not isolated but it is in contact with external heat baths, then, in some cases, the existence of a stationary measure is known \cite{EPR99, EH01}  although its properties are still not well understood.

Another possibility is to consider hard core potential, e.g. billiards. This is also a promising line of thought, very close in spirit to the one presented here. See section \ref{sec:hard} for details.

In the last years there have also been attempts to investigate models with mass transport, but with independent particles that can exchange energy only interacting with some array of localised systems, typically discs. Again such systems are in contact with reservoirs that can emit and absorb particles. This are intriguing and illuminating models for which is, at times, possible to establish the existence of a stationary measure and some of its properties \cite{EY04, EY06, CE09, Ya14}.  

In fact, there are many other relevant papers strictly connected to the matter at hand. It is impossible to quote them all, here is a very partial selection \cite{DL08, DKL08, BK07,  LZ10, Rue12}.

\section{A two steps strategy}
By the above discussion, the purely deterministic case seems completely out of reach of current techniques.  It is thus necessary to try to devise a line of attack that deals with the problems one at a time. The first, natural, idea is to leverage on our understanding of the dynamics for $\ve=0$. Of course, when $\ve=0$ there is no exchange of energy, so we must start to look at the case when $\ve$ is ``infinitesimally" small. One way to formalise precisely such a situation is to investigate if some universal behaviour takes place for small $\ve$.

\subsection{Soft core potentials}\label{sec:soft}\ \newline
This is the model we have discussed so far in the case $\nu=0$. Only, now we define the random variables $\cE_{L,\ve,x}(t)=\cm_x(\ve^{-2}t)$ and consider the limit $\ve\to 0$, keeping fixed the size of the system. In other words we look at the local energy when the interaction between nearby systems is very small, but rescale time in order to be able to see some evolution. 

The choice of the scaling $\ve^{-2}$ is due to the fact that, in equilibrium, the currents have zero average, hence we expect the exchange of energy between near by systems to be due to fluctuations. This means that, very naively, the variation of energy at site $x$ and time $t$ can be thought as the sum of  $t$ zero average independent random variables of size $\ve$. By the central limit theorem one then expects that a change of energy of order one takes place only  at time $\ve^{-2}$.

The above super naive picture can indeed be made rigorous in the special case of {\em contact} Anosov flows. Indeed, it is known that contact Anosov flows exhibit exponential decay of correlations \cite{Li04}. In the Hamiltonian \eqref{eq:ham-tot}, this corresponds to the requirement that the local Hamiltonian \eqref{eq:localh} be of the form $h(q,p)=\frac 12 \langle p,p\rangle$ and that $M$ is a compact manifold of strictly negative curvature. Note that the requirement that the local dynamics be a geodesic flow in negative curvature is not so artificial as it might appear at first sight. Indeed there exists mechanical models for which this is exactly the case \cite{HMc}. We have the following result.

\begin{thm}[\cite{DL11}]\label{thm:dolgo}   For each $L\in\bN$ and $n_*\geq 3$,\footnote{ The Theorem should also be true for $n_*=2$, but it is harder to prove. Instead it does not make sense for $n_*=1$, since in such a case the local Hamiltonian is completely integrable and the flow cannot be Anosov.} the process $\{\cE_{L,\ve,x}(t)\}_{x\in\Lambda_L}$, with initial conditions satisfying $\{\cE_{L,\ve,x}(0)=E_x>0\}_{x\in\Lambda_L}$, converges in law to a limit $\{\cE_{L,x}\}_{x\in\Lambda_L}$ satisfying the mesoscopic SDE
\begin{equation}\label{eq:dolgo}
\begin{split}
&d\cE_{L,x}=\sum_{|x-y|=1}\ba(\cE_{L,x},\cE_{L,y}) dt+\sum_{|x-y|=1}\bb(\cE_{L,x},\cE_{L,y}) dB_{x,y}\\
&\cE_{L,x}(0)=\bar\ce_x>0
\end{split}
\end{equation}
where  $\ba(\cE_{L,x},\cE_{L,y})=-\ba(\cE_{L,y},\cE_{L,x})$,  $\bb(\cE_{L,x},\cE_{L,y})=\bb(\cE_{L,y},\cE_{L,x})$ and $B_{x,y}=-B_{y,x}$ are independent standard Brownian motions. 
\end{thm}

The result includes the fact that the SDE is well posed, in the sense of the uniqueness of the martingale problem, \cite{SV06}.
To prove the latter it is necessary to show that zero is unreachable. Indeed, if zero were reacheable, then the equation \eqref{eq:dolgo} wold have to be supplemented by boundary conditions, since by definition energies are positive. In turn, to prove unreachability of zero it is necessary to acquire precise informations on the form of the diffusion coefficient and drift. In \cite{DL11} it is shown that
 $\ba,\bb\in\cC^\infty((0,\infty)^2)$ and, for $\cE_{L,x}\leq \cE_{L,y}$,
\[
\begin{split}
&\bb(\cE_{L,x},\cE_{L,y})^2=\frac{ A\cE_{L,x}}{\sqrt{2\cE_{L,y}}}+\cO\left(\cE_{L,x}^{\frac 32}\cE_{L,y}^{-1}\right)\\
&\ba (\cE_{L,x},\cE_{L,y}) =\frac{A n_*}{2\sqrt{2\cE_{L,y}}}+\cO\left(\cE_{L,x}^\frac 12\cE_{L,y}^{-1}\right),
\end{split}
\]
Note that the only invariant measures for \eqref{eq:dolgo} are measures absolutely continuous w.r.t. Lebesgue with density $h_\beta=\prod_{x\in\Lambda_L} \cE_{L,x}^{\frac{n_*}2-1} e^{-\beta\cE_{L,x}}$.

The SDE corresponds to a parabolic PDE with generator
\begin{equation}\label{eq:gen-meso}
\bX_L=\frac 1{2h_0}\sum_{|x-y|=1}(\partial_{\cE_{L,x}}-\partial_{\cE_{L,y}})h_0\ba^2(\partial_{\cE_{L,x}}-\partial_{\cE_{L,y}}).
\end{equation}
We have thus a {\em mesoscopic} equations in which the evolution of all the degree of freedom, apart from the energies, can be ignored. Even more remarkably, the generator $\bX_L$, with respect to the invariant measure, is {\em reversible}. 

This is a consequence of the microscopic reversibility of the flow. Indeed, the involution $(q,p)\to (q,-p)$ that exchanges the direction of time, reduces to the identity in the energy variables. Thus, we see not only how irreversibility arises (equations \eqref{eq:dolgo} are irreversible in the sense that they display a time arrow: a distribution tends to equilibrium going forward in time, but not going backward), but we see also a non trivial relation between (deterministic) reversibility for the microscopic dynamics and (stochastic) reversibility for the macroscopic one (which shows up already at the mesoscopic level).

The generator \eqref{eq:gen-meso} is a dynamics only on the energies, so it has the same flavour as $S$ in \eqref{eq:generats}. However, the associated current is not of gradient type. Yet, it is conceivable that the study of the dynamics \eqref{eq:dolgo} is much easier than the study of the original deterministic dynamics. So it natural to try to perform the hydrodynamics limit on the mesoscopic dynamics. 

To this end, as  we have already mentioned, it seems necessary to have a spectral gap of size $L^{-2}$ for the operator $\bX_L$.
 At the moment it is unclear if such a fact holds true or not, the problem stemming from the fact that at high energy the diffusion coefficient vanishes. This is a consequence of the fact that at high energy near by systems (in the original deterministic system) interact very little since the size of the potential is very small compared with the available energy.

The situation improves if one starts from a system with some stochasticity as in \cite{LO12}. Indeed, in \cite{LO12} is considered an anharmonic chain with and energy preserving noise and we establish the same type of result as in \eqref{eq:dolgo} but now
\[
\begin{split}
&\bb(\cE_{L,x},\cE_{L,y})^2\sim A\cE_x\cE_{L,y}\\
&\ba (\cE_{L,x},\cE_{L,y}) \sim \cE_{L,x}-\cE_{L,y}.
\end{split}
\]
For such a $\ba$ the needed spectral gap has been established by M. Sasada \cite{OS13} comparing $\bX_L$ with the generator of the Kac model. This suffices to prove that the fluctuations in equilibrium satisfy the heat equation \cite{LOS}.

Hence there is a concrete hope to obtain the heat equation starting from a deterministic dynamics via a two step procedure: first take a weak coupling limit to obtain a mesoscopic equation involving only the energies, then perform the hydrodynamic limit on the latter dynamics.

This is encouraging, yet a natural question arises: is there any relation between the behaviour of the original model, possible for $\ve$ very small, and the result of this two step procedure? To answer precisely to such a question would be equivalent to solve our original problem, however even an heuristic answer is not obvious.

A (non trivial) formal computation, see \cite{BHLLO}, shows that if $\kappa_\ve$ is the diffusivity, as defined in \eqref{eq:gk1}, for the original model (that we assume finite) and $\kappa_M$ the diffusivity of the mesoscopic  dynamics (that can be proven finite), then
\[
\begin{split}
&\kappa_\ve=\ve^2\kappa_M+\cO(\ve^3)\\
& \kappa_M =  \bE_0( {\bb}(\cE_0, \cE_1)^2) +  \sum_{x }\int_0^\infty \bE_0\left(  {\ba}(\cE_0,\cE_1)(0){\ba}(\cE_x,\cE_{x+1}) (s)\right)   ds.
\end{split}
\]
This suggests that the mesoscopic dynamics captures the main effect of the energy transport and that it is an effective approximation of the behaviour of the microscopic deterministic dynamics.

If the above were true, then it should be possible to use the stochastic dynamics as a first approximation of the long term statistical properties of the original dynamics well beyond the time scale $\ve^{-2}$, which is the time scale established by Theorem \ref{thm:dolgo}.
This is an intriguing possibility that leads to a rather vast research program.

\subsection{Hard core potential}\label{sec:hard}
Before continuing the discussion on the possibility to extend Theorem \ref{thm:dolgo}, it is worth to discuss a different possibility: hard core interactions.
Indeed, it is quite possible that the presence of hard core at the microscopic level does manifest itself also at macroscopic level. For example it is likely that hard core interactions do not manifest the property of a decreasing diffusion coefficient at high energies since when there is a collision the velocities change dramatically also at high energies, contrary to the case of soft interactions. 

Unfortunately, while hard core interactions may cure a problem they come at a high cost, since in such a case the discontinuity of the dynamics creates formidable technical problems. Yet, it is certainly very important and instructive to investigate this alternative.

\begin{figure}
\includegraphics[width=4.5cm,height=4.5cm]{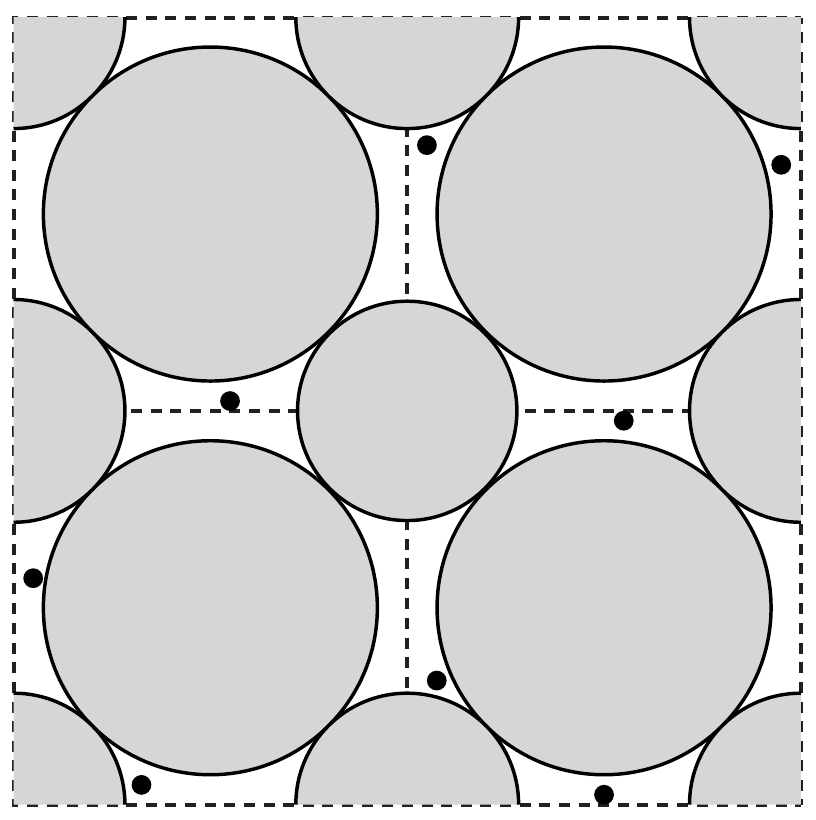}
\caption{Obstacles gray, particles black. Non interacting particles}\label{fig:1}
\end{figure}

\begin{figure}
\includegraphics[width=4.5cm,height=4.5cm]{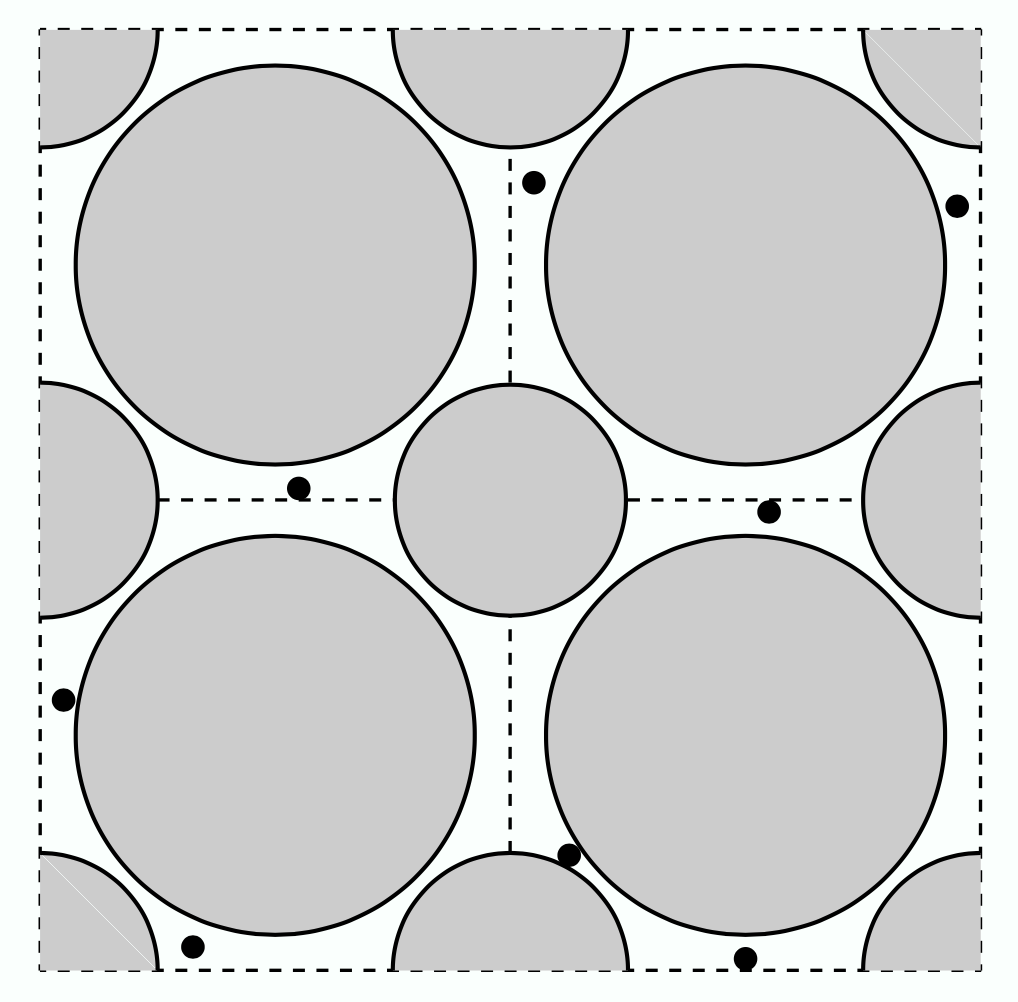}
\caption{Obstacles gray, particles black. Interacting particles}\label{fig:2}
\end{figure}

In this case, the unperturbed systems (corresponding to the Hamiltonian $H_{0,L}$) would consist of a region of size $L$ filled by disjoint billiards domains each containing a ball that can move freely a part for the elastic collision with the walls. In such a case the kinetic energy of each ball is conserved and there is no transport of mass or of energy, see figure \ref{fig:1}.
To perturb the system, instead of introducing a potential, we shrink a bit the obstacles, so that channels appear between the different tables. If the channels are large enough to allow near by particles to collide, but not so large as to allow the particles to escape the region in which they are confined, then we obtain a systems in which mass transport is still impossible, but energy transport is allowed, see figure \ref{fig:2}.

Some systems of these type are known to be ergodic \cite{BLPS} but, unfortunately, nothing is known about their mixing rate. In particular it is unknown if the Green-Kubo formula is well defined. Yet, one can imitate what has been done in the previous section: consider the limiting case in which the interaction are extremely rare and rescale the time so that, in average, a particle has one collision with another particle in a (macroscopic) unit time. This leads, again, to a two step route to the heat equation. This research program has been put forward in \cite{GG08} where the authors heuristically derive a mesoscopic equation describing the evolution of the energy with generator
\begin{equation}\label{eq:makiko}
\bX_L f(\ce)=\frac{1}{2}\sum_{x \in \Lambda_L} \sum_{\substack{y \in \Lambda_L \\ \|x-y\|=1}} \int_{-\pi}^{\pi} [ f(R^{xy}_{\theta}\ce)-f(\ce)] \rho(\theta) d\theta
\end{equation}
where $R^{xy}_{\theta}, x\neq y$ is a clockwise rotation of angle $\theta$ in the plane $(\ce_x,\ce_y)$. 
The generator \eqref{eq:makiko} is the analogous of \eqref{eq:gen-meso} and describe a jump process. This generator is know to have a spectral gap of order $L^{-2}$, \cite{GKS, Sa15}. The situation seems then very promising, unfortunately all attempts to derive rigorously \eqref{eq:makiko} have so far failed.
Nevertheless, lately there has been some notable technical progresses \cite{BDL, DN17,BNST17} and some relevant results on related models have appeared \cite{BGNST, DN16}.

\section{Partially hyperbolic Fast-slow systems and limit theorems}
At the end of section \ref{sec:soft} we came to the conclusion that \eqref{eq:dolgo} might hold for much longer times than $\ve^{-2}$ and that this, if true, could help in establishing the Green-Kubo formula and, ultimately, the heat equation. However, to investigate such a possibility is a non trivial task. A task that is best accomplished proceeding by intermediate steps. This leads us to the general problem of studying the long time validity of limit theorems in partially hyperbolic fast-slow systems. 

Fast-Slow systems are system in which there are two group of variables that evolve accordingly to very different time scales. For example, the Hamiltonian \eqref{eq:ham-tot}, with $M$ a compact Riemannian manifold in negative curvature and the local Hamiltonian \eqref{eq:localh} of the form $h(q,p)=\frac 12 \langle p,p\rangle$, yields a dynamics in which the variables $q_x, v_x=(2\cm_x)^{-1/2} p_x$ vary on a microscopic time scale of order one, while the variable $\cm_x=\frac 12 p_x^2$ varies on a timescale of order $\ve^{-2}$. Partial hyperbolicity stems form the fact that for $\ve=0$ the system foliates in uniformly hyperbolic systems, so the dynamics in the central direction is simply identity, and the central direction persists under perturbations \cite{HPS}. 

If we want to understand the behaviour of such systems for arbitrarily long times it is best to start form the simplest possible example.

\subsection{ The not so simple simplest example}\ \newline
We start by defining the one parameter family of maps $F_\ve\in\cC^4(\bT^2,\bT^2)$
\begin{equation}\label{eq:simpled}
F_\ve(x,\slow)=(f(x,\slow),\slow+\ve\omega(x,\slow)),
\end{equation}
where $\partial_x f(x,\slow)\geq \lambda >1$  is an expanding map for all $\slow$.
We then consider the dynamics $(x_n,\slow_n)=F_\ve^n(x_0,\slow_0)$ with initial conditions
\begin{equation}\label{eq:simpleic}
\bE(g(x_0,\slow_0))=\int_{\bT^1} \rho(x) g(x,\bar\slow_0) dx\;,
\end{equation}
where $\bar \slow_0\in\bT^1$, while $\rho\in\cC^2(\bT^1,\bR_+)$.

Let us compare this super simplified model with the Hamiltonian system \eqref{eq:ham-tot}. First of all, \eqref{eq:simpled} is in discrete time and not continuous time. This is technically much simpler, but morally not so different. 

More serious is the fact that the system is not time reversible and has no Hamiltonian or symplectic structure. This makes it rather artificial, however a time reversible system with some symplectic structure could be constructed using an Anosov map on the two torus instead of an expanding map of the circle. Hence, we can consider our model as a preliminary step toward a more realistic one. 

Next, \eqref{eq:simpled} has only two variables hence the question of taking the hydrodynamics limit makes no sense.\footnote{ Nonetheless one can consider many of such systems weakly coupled, whereby reproducing a situation in which it is possible to perform the hydrodynamics limit.  In the simple case in which the fast dynamics does not depend form the slow one, this has been done, obtaining indeed the heat equation, \cite{BK13}.} However, this model is intended only to explore the possibility to control the statistical properties of \eqref{eq:ham-tot} for a time longer than $\ve^{-2}$. Of course, ultimately this must be done with some uniformity on the number of degree of freedom, but if one cannot do it with one degree of freedom it does not make much sense to think about large systems.

On the bright side, \eqref{eq:simpled} has a conserved quantity ($\slow$, which plays the role of the local energy in \eqref{eq:ham-tot}) for $\ve=0$ and, for $\ve$ small is a fast-slow system. The quantity $\ve\omega$, which determines the change of the almost conserved quantity $\slow$, plays the role of the current. 

The local dynamics depends on the conserved quantity $\slow$ as the local hamiltonian dynamics in \eqref{eq:ham-tot} depends on the local energy. The local dynamics $f(\cdot, \slow)$ have a strong chaotic character similar to the hypothesis that the local, unperturbed, hamiltonian flows is a contact Anosov flow (or, more generally, enjoys exponential decay of correlations), hence the partial hyperbolicity. The initial conditions are very similar as one can fix exactly the almost conserved quantity (slow variable) but must have a smooth distribution for the fast variables, similarly to \eqref{eq:inco}.

It is well known, \cite{Babook}, that our hypotheses on \eqref{eq:simpled} imply that
\begin{enumerate}
\item for each $z\in\bT^1$, $f(\cdot, z)$ has a unique SRB measure $\mu_\slow$ which is absolutely continuous with respect to Lebesgue and has density $h(\cdot,\slow)$
\item $h\in\cC^2(\bT^2,\bR_+)$.
\end{enumerate}
The above facts take care of another apparent difference between \eqref{eq:ham-tot} and \eqref{eq:simpled}: the former has an explicit and natural invariant measure (Liouville). Now we know that, even though not totally apparent, the same holds for \eqref{eq:simpled}. The measures $\mu_\slow$ will play the role of the equilibrium measures. 

However, there is a last issue: due to the reversibility and the hamiltonian structure in \eqref{eq:ham-tot} the average of the current is always zero. This is the reason why the evolution of the energy happens on the time scale $\ve^{-2}$ rather than on the scale $\ve^{-1}$. In the present simplified setting this would correspond to the conditions $\bar\omega(\slow):=\mu_\slow(\omega(\cdot, \slow))=0$. Unfortunately, this turns out to be a much harder problem. At the moment are available partially satisfactory results only in the case $\bar\omega(\slow)\neq 0$. More precisely, in the case in which $\bar\omega$ has only finitely many non-degenerate zeroes. In this case the natural time scale in which the slow variable evolves is $\ve^{-1}$, however the problem of understanding the statistical properties of the system for arbitrarily long times remain a non trivial challenge and its study is a preliminary step to attack the, harder, case $\bar\omega\equiv 0$.

To describe the existing results it is convenient to introduce the continuous paths
\[
\slow_\ve(t)=\slow_{\lfloor\ve^{-1} t\rfloor}+(\ve^{-1} t- \lfloor\ve^{-1} t\rfloor)(\slow_{\lfloor\ve^{-1} t\rfloor+1}-\slow_{\lfloor\ve^{-1} t\rfloor}),\quad t\in[0,T].
\]
The paths $\slow_\ve\in\cC^0([0,T],\bR)$ are random variables due to the randomness of the initial conditions.
Since the $\{\slow_\ve\}$ are uniformly  Lipschitz, they belong to a compact set in $\cC^0([0,T],\bR)$, hence they have convergent subsequences. It is possible to show that  all the accumulation points $\oTheta$ must satisfy the ODE
\[
\begin{split}
&\dot\oTheta=\bar\omega(\oTheta)\\
&\oTheta(0)=\bar\slow_0\\
&\bar\omega(z)=\int_{\bT^1}\omega(x,z) h(x,z) dx=\mu_\slow(\omega(\cdot,\slow)).
\end{split}
\]
This type of results goes back, at least, to Anosov \cite{An60} and Bogolyubov-Mitropolskii \cite{BoM61} in the early '60.

Next, let us consider the quantity $\zeta_\ve(t)=\ve^{-\frac 12}\left[\slow_\ve(t)-\oTheta(t)\right]$. These are the fluctuations around the average. To discuss this case we need to recall that a function $\phi\in\cC^0(\bT)$ is said to be a
\emph{(continuous) coboundary (with respect to a map $f:\bT\to\bT$)} if there exists
$\beta\in\cC^0(\bT)$ so that
\begin{align*}
\phi=\beta-\beta\circ f.
\end{align*}
Two functions $\phi_1,\phi_2\in\cC^0(\bT)$ are said to be \emph{cohomologous (with respect
  to $f$)} if their difference $\phi_2-\phi_1$ is a coboundary (with respect to $f$).  We make the
non-degeneracy assumption that, for each $\slow\in\bT$, the function $\omega(\cdot,\slow)$ is not
  cohomologous to a constant  with respect to $f_\slow$. Note that in~\cite{DeL1} it is shown that
  this assumption is in fact generic in $\cC^2$. 

A computation using the decay of correlations of the maps $f(\cdot,\slow)$ yields
\[
\bE([\zeta_\ve(t)-\zeta_\ve(s)]^4)\leq C |t-s|^2.
\]
Hence, by Kolmogorov criteria, the sequence is tight.
It is possible to show that the accumulation points $\zeta$ of $\zeta_\ve$ satisfy
\begin{equation}\label{eq:dolgo-av}
\begin{split}
&d\zeta=\bar\omega'(\oTheta(t))\zeta(t) dt+ \bVar(\oTheta(t)) dB\\
&\zeta(0)=0
\end{split}
\end{equation}
where $B$ is the standard Brownian, $\bVar>0$ is given by the Green-Kubo formula
\[
\begin{split}
  \bVar(\slow)^2 =& \mu_ \slow\left(\hat\omega(\cdot,\slow)\hat\omega(\cdot,\slow)\right)+ 2\sum_{m=1}^{\infty}
   \mu_ \slow\left( \hat\omega(f_\slow^m(\cdot), \slow)\hat\omega(\cdot,\slow)\right),
\end{split}
\]
and we have used the notation $f_\slow(x)=f(x,\slow)$.

As the above equation has a unique solution, this identifies the limit.

This type of results are much more recent and, in the above form, have been obtained by Dolgopyat and Kifer at the beginning of the new millenium, \cite{Do05, Kifer04}, but see \cite{DeL15} for a pedagogical exposition.

We have thus seen that $\slow_\ve$ is close to $\oTheta+\sqrt\ve \zeta$. 
On the other hand it is possible to show, \cite{Ki76}, that $\oTheta+\sqrt\ve \zeta$ is close to the solution $\sTheta$ of the stochastic differential equation
\begin{equation}\label{eq:wfeq}
\begin{split}
&d\sTheta=\bar\omega(\sTheta)dt+\sqrt\ve  \bVar(\sTheta)dB \\
&\sTheta(0)=\bar \slow_0.
\end{split}
\end{equation}
Thus the motion is described by an ODE with a small random noise of the type introduced by Hasselmann \cite{H76} to model climate and extensively studied by Wentzell--Freidlin \cite{FW69, FW79} and Kifer \cite{Ki74, Ki77, Ki81} in the 70's.

The above is the equivalent, in the present context, of Theorem \ref{thm:dolgo}. We can now pose for the current model the question that we would like to answer in the previously described context: what happens on time scales longer than $\ve^{-1}$. A first result is the following:
\begin{thm}[\cite{DLPV} Corollary 3.3] \label{cor:third-step}
For any $\beta>0$, $\alpha\in(0,\beta)$, $\ve\in (0,\ve_0)$, and $t\in[\ve^{1/2000},\ve^{-\alpha}]$, there exists $C_\beta>0$ and a coupling $\bP_c$ between $\slow_\ve(t)$ and $\sTheta(t)$, such that:
\[
\bP_c(|\slow_\ve-\sTheta(t)|\geq \ve)\leq C_\beta \ve^{1/2-\beta}.
\]
\end{thm}
The above result is based on a drastic sharpening of \eqref{eq:dolgo-av}, which amounts to a local central limit theorem with error term for the diffusion limit.

\begin{thm}[{\cite[Theorem 2.7]{DeL1}}]\label{thm:lclt}
 For any $T>0$, there exists $\ve_0>0$ so that the following holds.  For any $\beta>0$,
    compact interval $I\subset\bR$, $|I|\leq 1$, real numbers $\shiftPar>0$, $\ve\in(0,\ve_0)$,
    $t\in[\ve^{1/2000},T]$,  we have:
  \[
 \frac{ \bE(\zeta_\ve(t,\cdot)\in
    \veh I + \shiftPar)}{\sqrt\ve}= \Leb\, I\left[
      \frac{e^{-\shiftPar^2/2\Var_t^2(\bar \slow_0)}}{\Var_t(\bar\slow_0)\sqrt{2\pi}}\right]+\cO(\ve^{\efrac 12-\beta}).
  \]
  where the variance $\Var_t^2(\slow)$  reads
  \[
  \Var_t^2 (\slow)= \int_0^t e^{2\int_s^t\bar\omega'(\oTheta(r,\slow))\deh r}\bVar^2(\oTheta(s,\slow))\deh s.
  \]
\end{thm}

Theorem \ref{cor:third-step} says that the deterministic dynamics remains close to the stochastic one for a time of order almost $\ve^{-\frac 32}$. Since \eqref{eq:wfeq} reaches at lest a metastable state in a time of order $\ve^{-1}\ln\ve^{-1}$, \cite{FW79}, it follows that also the deterministic system must reach similar states.
This control should be sufficient to start an investigation of the even longer time properties of the system. 

It turns out that current techniques to study the statistical properties of partially hyperbolic systems depend on the positivity or negativity of the central Lyapunov exponent. So at the moment it is unclear in which generality the program can be completed. 
Here we present the best available result, but before stating it  is necessary to introduce some notation.

We say that a Lipschitz path $\fpath$ of length $T$ is \emph{admissible} if for any $s\in[0,T]$,
$\dfpath(s)\subset\intr\Omega(h(s))$,\footnote{ For each $s\in[0,T]$, $\dfpath(s)$ is the \emph{Clarke generalized derivative} of
$\fpath$ as the set-valued function:
\begin{align*}
  \dfpath(s) = \text{hull}\{\lim_{k\to\infty} \fpath'(s_k) \st s_k\to s  \}.
\end{align*}
The set $\dfpath(s)$ is compact and non-empty (see~\cite[Proposition~2.1.5]{Clarke}) and so is its graph.}
 where, for $z\in\bT$, we define the (non-empty, convex and compact) set
\[
  \Omega(\slow)=\{\mu(\omega(\cdot,\slow))\,| \,\mu\text{ is a
  }f_\slow\text{-invariant probability}\}.
\]
The last condition is:
\begin{itemize}
  \item there exists $i\in\{1,\cdots,\nz\}$ so that for any
  $\slow\in\bT$, there exists an \admissiblep{\slow}{\slow_{i,-}}{}.  We can always
  assume, without loss of generality, that $i=1$.
\end{itemize}
Observe that the above condition is trivially satisfied if $\nz=1$.

\begin{thm}[\cite{DeL2}Main Theorem]
The map $F_\ve$ admits a unique SRB
  measure $\mu_\ve$. This measure enjoys exponential decay of correlations for H\"older
  observables.  
  More precisely: there exist $C_1,C_2,C_3,C_4>0$ (independent of $\ve$) such that, for
  any $\alpha\in (0,3]$ and $\beta\in(0,1]$, any two functions
  $A\in\cC^\alpha(\bT^2)$ and $B\in \cC^\beta(\bT^2)$:
  \begin{align*}
    \left| \Leb(A\cdot B\circ F_\ve^n) - \Leb(A)\mu_\ve(B)\right|\leq
    C_1\sup_ \slow\|A(\cdot,  \slow)\|_{\alpha}\sup_x\|B(x,\cdot)\|_{\beta}
    e^{-{\alpha\beta} c_\ve n},
  \end{align*}
  where
  \[
    c_\ve=
    \begin{cases}
      C_2\ve/\log\vei&\text{if }\nz=1,\\
      C_3\exp(-C_4\vei)&\text{otherwise.}
    \end{cases}
  \]
\end{thm}

The proof of the above results is based on the standard pair technique \cite{Do05} and Theorem \ref{thm:lclt}, but also on a considerable sharpening of the large deviation results previously obtained for uniformly hyperbolic dynamical systems, notably \cite{Kifer09}, see \cite{DeL1} for details.

Also note the extremely slow decay of correlations in the case in which more than one sink is present. This is a well known phenomena: {\em metastability}. A phenomena widely studied in stochastic equations like \eqref{eq:wfeq}, see \cite{FW79, OV05}, but seen here for the first time in a purely deterministic setting.

\section{ Final considerations}
The previous sections show on the one hand that even establishing partial results for a very simplified system entails a tremendous amount of work. On the other hand we have shown that the research program of studying the dynamics of \eqref{eq:ham-tot} is not totally hopeless, especially if the arguments could be substantially simplified.

In particular, we have put forward the general philosophy of proving that a deterministic dynamics behaves like a stochastic one for a time long enough for the stochastic dynamics to exhibit some asymptotic property. This allows to deduce that the deterministic system shares such  asymptotic behaviour. This approach seems to be a powerful point of view that can be applied to many other contexts.

\end{document}